# The qualitative accuracy of the Becker-DeGroot-Marshak method

## Maximilian Späth[*]

University of Potsdam, Department of Economics, Potsdam, Germany

Email: maximilian.spaeth@uni-potsdam.de

February 08, 2023

**Abstract**

*The Becker-DeGroot-Marshak method is widely used to elicit the valuation that an individual assigns to an object. Theoretically, the second-price structure of the method gives individuals the incentive to state their true valuation. Yet, the elicitation method's empirical accuracy is subject to debate. With this paper, I provide a clear verification of the qualitative accuracy of the method. Participants of an incentivized laboratory experiment can sell a virtual object. The value of the object is publicly known and experimentally varied in a between-subjects design. Replicating previous findings on the low quantitative accuracy, I observe a very small share of individuals placing a payoff-optimal stated valuation. However, the analysis shows that the stated valuation increases with the value of the object. This result shows the qualitative accuracy of the BDM-method and suggests that the method can be applied in comparative studies.*

---

[*]I have received helpful comments from participants of the annual conference of the Gesellschaft für experimentelle Wirtschaftsforschung (GfeW) in Magdeburg. In particular, I would like to thank Lisa Bruttel and her team, Claudia Keser, Michael Seebauer, Max R. P. Grossmann und Susanna Grundmann. Declarations of interest: none.

## 1. Introduction

The Becker-DeGroot-Marshak (1964, henceforth BDM) method is an often-used elicitation strategy to measure the valuation that individuals assign to an object.[1] In the version used by Cason and Plott (2014), individuals take part in second-price selling auction. Individuals are endowed with an object and submit a minimum selling price. If this offer price exceeds a random price, the object is sold at a random price. Otherwise, the object stays with the individual. The second-price structure gives the incentive to state an offer price equal to the value of the object. Hence, from a theoretical perspective, the BDM-method can be seen as a valid though not undisputed strategy to elicit the true valuation.[2]

Far more debatable is how the method can be applied in actual empirical research. The seminal paper by Cason and Plott (2014) and the replication by Bull et al. (2019) show a low quantitative accuracy of the BDM-method, in the sense that the method mostly fails to elicit the exact valuation that individuals place on object. The principal idea of the authors is to elicit the valuation for an object of known value. Cason and Plott (2014) argue that the valuation for an object worth $2 would most probably be $2. Yet, Cason and Plott (2014) as well as Bull et al. (2019) find that only a small share of individuals (16.7 percent and 7.9 percent, respectively) state this amount in a BDM-setup.[3] Further emerging literature finds significant differences between the valuation values elicited with the BDM-methods and those elicited with other incentive compatible measures (Noussair et al., 2004; Lusk and Schroeder, 2006; Blavatskyy and Köhler, 2009; Flynn et al., 2016; Freeman et al. 2016, Hamukwala et al., 2019; Dinc-Cavlak and Ozdemir, 2021) or differences between variants of the BDM-mechanism (James, 2007; Blavatskyy and Köhler, 2009).

Given the reported low quantitative accuracy, the aim of this paper is to understand whether the BDM-method is a qualitatively accurate elicitation tool. In other words, the contribution of this paper is to experimentally check whether the elicited valuation for an object increases with the value of the object. Building on the design introduced by Cason and Plott 2014, subjects of a laboratory experiment have the option to sell a virtual object of known value. In a between-subjects design, the monetary value of the object is experimentally varied. This provides a clear test of the qualitative accuracy of the method.

---
[1] The original article by Becker et al. (1964) is cited at least 3400 times by February 2023 (according to Google Scholar).
[2] Claims that the BDM-method does not even theoretically reveal the true willingness to pay are made by Karni and Safra (1987) and Horowitz (2006), among others.
[3] In contrast, Burchardi et al (2021) observe a high share of rational bids of around 86 percent in a lab-in-the-field experiment.

The results show a high qualitative accuracy of the BDM-method. The stated valuations increase with the objective value of the good. However, the quantitative accuracy of the method is low. As Cason and Plott (2014) and Bull et al. (2019), I find that only a small share of individuals states a payoff-maximizing value. Instead, subjects systematically overstate their valuation. These findings imply that the BDM-mechanism with only minimalistic instructions is not suitable to elicit the valuation for single objects. Yet, the mechanism can be recommended for comparative studies. Even with minimalistic instructions, the method can capture valuation differences between objects.

## 2. Experimental Design

The core of the experimental design is the second-price selling auction as used by Cason and Plott (2014). The instructions can be found in Fig 1. Instructions are minimalistic and without examples. The German instructions can be found in the Appendix. Subjects receive an object M. They have the opportunity to sell the object. Subjects enter an amount of Experimental Currency Units (ECU) at which they want to offer the object. In line with the BDM-approach, this offer is compared to a random amount. The amount is taken from the interval between 0 ECU and 100 ECU, where each integer value is drawn with equal probability. If the price entered by the subject is below or equal to the random price, the subject will sell the object. Otherwise, the subject keeps the object.

Importantly, the object M has a known economic value to the subject. If the subject keeps the object (in other words, does not sell it), the subject will receive a pre-defined value from the experimenter. This value is 60 ECU in treatment HIGH, while it is 30 ECU in LOW. Allocation of subjects into treatments is random. Given the structure of the game, stated valuations of 60 ECU and 61 ECU respectively 30 ECU and 31 ECU are payoff-maximizing. In line with Cason and Plott (2014) and Bull et al. (2019), I argue that the true valuation would be equal to this payoff-maximizing amount.

> You now have the possibility to earn additional ECU.
>
> You receive a virtual object M. If you keep the object, you will receive 60 [30] ECU.
>
> However, you have the opportunity to sell the object.
> For this, you enter a price at which you would like to offer the object.
>
> The actual selling price is a randomly drawn amount between 0 ECU and 100 ECU.
> Each integer price between 0 ECU and 100 ECU is possible with the same probability.
>
> If the stated price is **below or equal** to the random selling price, then you will sell the object at the random selling price.
> If the stated price is **above** the random selling, then you won't sell the object. In this case, you will receive 60 [30] ECU for the object.
>
> At which price (in ECU) do you want to offer the object?

**Fig 1** Instructions of the BDM-mechanism with both values of the object (variation in square brackets). Translated from German into English.

The experiment is incentivized and pre-registered on OSF (Späth: The qualitative and quantitative accuracy of the Becker-DeGroot-Marshak method).[4] Every 10 ECU earned are exchanged to EUR 0.10 at the end of the experiment. Subjects are informed about the result of the selling auction directly after they confirmed their offer price. The BDM-experiment takes place after an unrelated experiment on investment decisions into lotteries. The pre-registration of the other experiment can be found on OSF (Späth and Goller: The grass is greener on which side again? Irrelevant information and the stickiness of reference risk choices).[5]

The experiment was conducted at the Potsdam Laboratory for Economic Experiments using zTree (Fischbacher, 2007) and ORSEE (Greiner, 2015). The experimental sessions took place in autumn 2022. As pre-registered, 150 independent observations were collected, with 75 observations in each treatment. Roughly 47 percent of the subjects state to be female. Randomization into treatments is imperfect in the sense that the share of women receiving the high value good is larger than the share of the joined group of other genders (here: male or diverse) who receive the high value good (Fisher's exact test: $p = 0.022$). The average reported age is 23 years. Subjects report on average four participations in laboratory experiments (including the current study). For age and experimental participations, the randomization worked as expected.[6]

---
[4] Link: https://osf.io/eucqg
[5] Link: https://osf.io/s8p9x/

## 3. Results

The main aim of this study is to analyze whether the BDM-mechanism is qualitatively accurate or not. High qualitative accuracy is given when the mechanism can capture changes in the known value of the object. In other words, the stated valuation of the object would increase with the pre-defined value of the object. To analyze the qualitative accuracy, I test the pre-registered Hypothesis 1 stating that the offer price in treatment HIGH is higher than the offer price in treatment LOW.

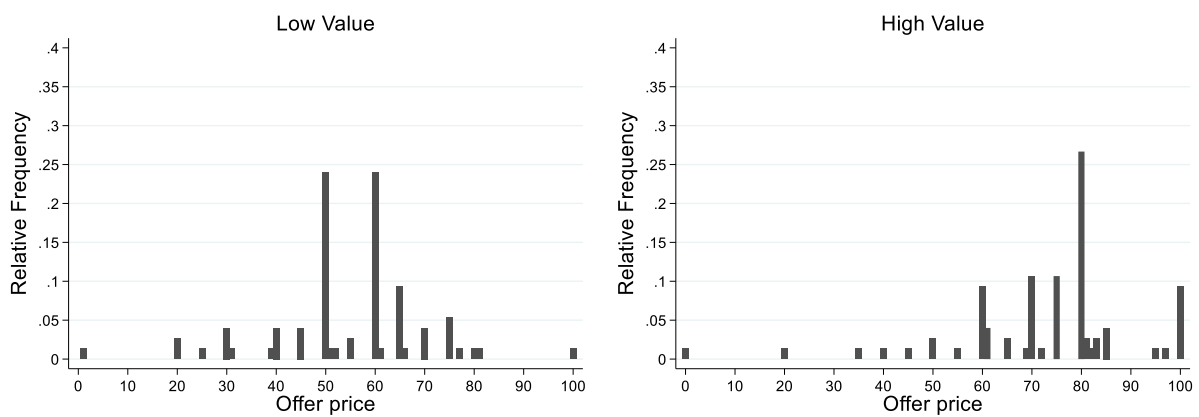

**Fig 2** Spikeplot on the distribution of offer prices per treatment (treatment LOW on the left, treatment HIGH on the right)

The analysis shows that the BDM-mechanism is indeed qualitatively accurate. In line with Hypothesis 1, I find higher offer prices in HIGH than in LOW. While the average offer price in treatment HIGH is 73.15 ECU (median: 75, sd: 17.51), I observe an average offer price of 54.27 ECU (median: 60, sd: 15.13) in treatment LOW. Fig 2 exhibits that 80 ECU is the most frequent offer price in HIGH. The two most prominent offer prices in LOW are 50 ECU and 60 ECU. I calculate an effect size d (Cohen, 1988) of d = 1.11. The difference in offer prices between treatments is statistically significant (Wilcoxon rank-sum test: $p < 0.000$). An ex-post analysis using G*power (Faul et al., 2007) shows an achieved power of larger than 0.99. Given the effect size and setting alpha to 0.05, a total sample size of 40 observations would be sufficient to achieve a power of at least 0.95. These findings highlight the large qualitative accuracy of the BDM-mechanism.

---
[6] Wilcoxon rank-sum tests: Age $p = 0.278$; experimental participations $p = 0.814$.

The second aim of the experiment is to test the quantitative accuracy of the BDM-mechanism. In line with Cason and Plott (2014), I expect to find that subjects state a valuation that is significantly above the payoff-maximizing value. The pre-registered Hypothesis 2 states the expectation that the offer price in HIGH is higher than 61 ECU and the offer price in LOW to be higher than 31 ECU.

Replicating the findings by Cason and Plott (2014) and Bull et al. (2019), I observe a low quantitative accuracy of the BDM-mechanism. Fig 2 shows that only 13.33 percent of the subjects (10 of 75) in HIGH choose an offer price equal to 60 ECU or 61 ECU. In LOW, only 5.33 percent of the subjects (4 of 75) choose an offer price equal 30 ECU or 31 ECU. In both treatments, most subjects choose an offer price above the payoff-maximizing value. The differences to the two respective payoff-maximizing references are statistically significant (Wilcoxon signed-rank tests: for both $p < 0.001$). In line with Hypothesis 2, I find that most subjects overstate their valuation of the object.

As pre-registered, I finally test which factors are influencing the accuracy of the BDM-auction. Table 1 shows the results of OLS-regressions. In Column (1) and Column (2), the outcome variable is the offer price. In Column (3) and Column (4), the outcome is whether the offer price is rational in the sense that is payoff-maximizing. I find that the stated gender, age, and number of participations in experiments have no significant impact, neither on the amount of the offer price nor on its rationality. In additional support of Hypothesis 1, the offer price is significantly larger in HIGH than in LOW ($p = 0.000$). Yet, the treatment variation does not impact the rationality of the offer price. Exploratorily, I add a variable on the reading time of the instruction of the BDM-mechanism. Column (2) shows that the reading time does not impact the amount of the offer price. Importantly, Column (4) exhibits that the rationality of the bid significantly increases with reading time of the instructions ($p = 0.015$). This positive correlation suggests that game-form misconceptions (Cason and Plott, 2014) are indeed a driver of the low quantitative accuracy of the BDM-mechanism.

**Table 1** Ordinary least-squares linear regression on the amount respectively the rationality of the offer price.

|  | (1) | (2) | (3) | (4) |
|---|---|---|---|---|
|  | Offer price amount | | Offer price rationality | |
| HIGH | 18.509*** (2.769) | 19.108*** (2.857) | 0.085 (0.049) | 0.055 (0.049) |
| Female | -2.295 (2.776) | -2.391 (2.781) | -0.030 (0.049) | -0.026 (0.048) |
| Age | -0.655 (0.415) | -0.630 (0.416) | -0.051 (0.007) | -0.006 (0.007) |
| Number experiments | 0.284 (0.494) | 0.221 (0.499) | 0.014 (0.009) | 0.017 (0.009) |
| Reading time |  | -0.042 (0.048) |  | 0.002* (<0.001) |
| Constant | 69.597*** (9.138) | 72.212*** (9.635) | 0.134 (0.161) | 0.006 (0.166) |
| N | 150 | 150 | 150 | 150 |
| R² | 0.257 | 0.261 | 0.039 | 0.078 |

Note: Standard errors in parentheses. Offer price amount in ECU. Offer price rationality has the value of if a payoff-maximizing offer price is chosen, otherwise zero. Reference category for treatment HIGH: treatment LOW. Reference category for female: jointly male and diverse. Age in years. Number experiments includes the current study. Reading time in seconds. * p < 0.05, ** p < 0.01, *** p < 0.001.

### 4. Conclusion

The aim of this research is to test the accuracy of the Becker-DeGroot-Marshak (1964, BDM) method to elicit the valuation that individuals assign to an object. Replicating the results reported by Cason and Plott (2014) and Bull et al. (2019), I find a low quantitative accuracy of the BDM-mechanism. Subjects report values that are significantly above a rational reference. Importantly, I observe a high qualitative accuracy of the BDM-method. The stated valuation increases with the objective value of the object.

This implies that researchers can use the BDM-method as a tool in comparative studies, in which the aim is to compare the valuations that individuals assign to several goods. Even with minimalistic instructions a valuation comparison between the goods is feasible. Yet, the findings on the low quantitative accuracy of the BDM-method highlight that minimalistic instructions might be insufficient in studies with the aim to elicit the exact valuation that individuals assign to a single good. A promising alternative proposed by Dizon-Ross and Jayachandran (2022) is the use of a benchmark good. Besides, the positive correlation of the rationality of stated valuations with the reading time of the instructions underlines the relevance of the literature on how to explain the BDM-auction to experimental participants (e.g., Bartling et al., 2015; Burfurd and Wilkening, 2018). The approach to conduct the BDM-mechanism with a good of known value can be used in pre-testing subjects' understanding of experimental BDM-instructions and be part of control questions in lab, online, or lab-in-the-field experiments.

**Statements and Declarations**

The author declares no competing interests. The research funded was funded by the chair of Economics, especially Markets, Competition & Institutions at the University of Potsdam. The study was approved by the German Association for Experimental Economic Research (Institutional Review Board Certificate No. IH8xynki).

**Appendix**

Instructions (German, original language)

Sie haben nun die Möglichkeit, weitere ECU zu verdienen.

Sie erhalten ein virtuelles Objekt M. Sofern Sie das Objekt behalten, erhalten Sie 60 [30] ECU.

Sie haben jedoch die Möglichkeit das Objekt zu verkaufen.
Sie geben dazu einen Preis ein, zu dem Sie das Objekt anbieten möchten.

Der tatsächliche Verkaufspreis ist ein zufällig gezogener Betrag zwischen 0 ECU und 100 ECU.
Jeder ganzzahlige Preis zwischen und inklusive 0 ECU und 100 ECU ist mit identischer Wahrscheinlichkeit möglich.

Ist der von Ihnen genannte Preis **kleiner als oder gleich** dem zufälligen Verkaufspreis, dann verkaufen Sie das Objekt zu dem zufälligen Verkaufspreis.
Ist der von Ihnen genannte Preis **größer** als der zufällige Verkaufspreis, dann verkaufen Sie das Objekt nicht. In diesem Fall erhalten Sie 60 [30] ECU für das Objekt.

Zu welchem Preis (in ECU) möchten Sie das Objekt anbieten?